\documentstyle[12pt,epsfig]{article}
\def\be{\begin{equation}} \def\ee{\end{equation}} \def\bea{\begin{eqnarray}}
\def\eea{\end{eqnarray}} \def\nnb{\nonumber}

\begin{document}

\hfill{February 20, 2025}

\begin{center}
\vskip 6mm 
\noindent
{\Large\bf  
	Fixing effective range parameters
	in elastic $\alpha$-$^{12}$C scattering: 
	an impact on resonant $2_4^+$ state of $^{16}$O 
	and $S_{E2}$ factor of $^{12}$C($\alpha$,$\gamma$)$^{16}$O 
}
\vskip 6mm 

\noindent
{\large 
Shung-Ichi Ando\footnote{mailto:sando@sunmoon.ac.kr}, 
}
\vskip 6mm
\noindent
{\large \it
Department of Display and Semiconductor Engineering, and \\
Research Center for Nano-Bio Science, 
Sunmoon University,
Asan, Chungnam 31460,
Republic of Korea
}
\end{center}

\vskip 6mm

Elastic $\alpha$-$^{12}$C scattering for $l=2$ and $E2$ transition of
radiative $\alpha$ capture on $^{12}$C, $^{12}$C($\alpha$,$\gamma$)$^{16}$O,
are studied in cluster effective field theory.
Due to the problem in fixing 
the asymptotic normalization coefficient (ANC) of the subthreshold $2_1^+$ 
state of $^{16}$O, equivalently, 
the effective range parameters of the $2_1^+$ state, 
from the elastic scattering data, we introduce
the conditions that lead to a large value of the ANC.
In addition, $d$-wave phase shift data of the elastic scattering
up to the $\alpha$ energy, $E_\alpha=10$~MeV, which contain resonant
$2_4^+$ state of $^{16}$O, are also introduced in the study.
Applying the conditions, 
the parameters of the $S$ matrix of the elastic scattering for $l=2$
are fitted to the phase shift data, and the fitted parameters are employed
in the calculation of astrophysical $S_{E2}$ factor of 
$^{12}$C($\alpha$,$\gamma$)$^{16}$O; we extrapolate the $S_{E2}$ factor
to the Gamow-peak energy, $E_G=0.3$~MeV.
We find that the conditions lead to the significant effects 
in the observables of the $2_4^+$ state of $^{16}$O
and the estimate of the $S_{E2}$ factor at $E_G$ and 
confirm that the ANC of the $2_1^+$ of $^{16}$O cannot be determined
by the phase shift data for $l=2$. 

\newpage 
\vskip 2mm \noindent
{\bf 1. Introduction}

The radiative $\alpha$ capture on $^{12}$C, $^{12}$C($\alpha$,$\gamma$)$^{16}$O,
is one of the fundamental reactions in nuclear astrophysics, which 
determines, along with the triple $\alpha$ reaction, the C/O ratio in the 
core of a helium-burning star~\cite{f-rmp84}.
It provides an initial condition for computer simulations of star 
evolution~\cite{ww-pr93,ietal-aj01} 
and leads to a significant influence on the results of star
explosions and nucleosynthesis~\cite{jj-20}. 
The reaction rate, or equivalently the astrophysical $S$ factor of 
$^{12}$C($\alpha$,$\gamma$)$^{16}$O at the Gamow-peak energy, $E_G=0.3$~MeV,
however, 
has not been measured in an experimental facility because of the Coulomb 
barrier. One needs to employ a theoretical model, fit the model parameters
to experimental data measured at a few MeV energy, and extrapolate 
the reaction rate to $E_G$. 
While it is known that $E1$ and $E2$ transitions of 
$^{12}$C($\alpha$,$\gamma$)$^{16}$O are dominant due to the subthreshold
$1_1^-$ and $2_1^+$ ($l^\pi_{ith}$) states of $^{16}$O, 
whose binding energies respected
to the $\alpha$-$^{12}$C breakup energy are $B_1=0.045$~MeV and $B_2=0.245$~MeV,
respectively~\cite{twc-npa93}. 
During the last half-century, many experimental and theoretical studies
on the reaction have been carried out. For a review, refer, e.g., to
Refs.~\cite{bb-npa06,cetal-epja15,bk-ppnp16,detal-rmp17,a-epja21}. 
(For a brief review, see Ref.~\cite{a-ha23}.)

We have been studying reactions related to 
$^{12}$C($\alpha$,$\gamma$)$^{16}$O by constructing a low-energy effective
field theory (EFT) based on the methodology of quantum field
theory~\cite{w-physica79,hkk-rmp20,dgh}. 
When constructing an EFT, 
one first chooses a typical scale of a reaction to study and
then introduces a large scale by which relevant degrees of freedom
at low energy are separated from irrelevant degrees of freedom from high
energy. 
We choose the Gamow-peak energy, $E_G=0.3$~MeV, as a typical energy scale;
a typical momentum scale would be $Q=\sqrt{2\mu E_G}=40$~MeV where
$\mu$ is the reduced mass of $\alpha$ and $^{12}$C.~\footnote{
	A typical length scale of the reaction is $Q^{-1} = 4.8$~fm.
} Because the typical wavelength of the reaction is larger than the size of
the nuclei, nucleons inside of the nuclei would be irrelevant; 
we assign $\alpha$ and $^{12}$C as structure-less (point-like)
spin-0 scalar fields. We then choose the energy difference between 
$p$-$^{15}$N and $\alpha$-$^{12}$C breakup energies of $^{16}$O;
$\Delta E = 12.13 - 7.16 = 4.97$~MeV, as the high energy (separation) scale;
the high momentum scale is $\Lambda_H = \sqrt{2\mu \Delta E} = 160$~MeV.
The theory provides us with a perturbative expansion scheme and the
expansion parameter would be $Q/\Lambda_H = 1/4$. 
The $p$-$^{15}$N system is now regarded as irrelevant degrees of freedom
and integrated out of the effective Lagrangian, whose effects are
embedded in the coefficients of terms of the Lagrangian. 
Those coefficients can, in principle, be determined from the mother theory,
while they, in practice, are fixed by using experimental data or empirical
values of them. 
Because of the perturbative expansion scheme of EFT, by truncating the
terms up to a given order, one can have an expression of reaction
amplitudes in terms of a few parameters for each of the reaction channels. 
This approach was recently used for the studies of reactions,
which are important in nuclear astrophysics,
such as elastic $p$-$^{12}$C scattering~\cite{ipsh-prc24}, 
elastic $d$-$\alpha$ scattering~\cite{nra-epja23},
and radiative proton capture on $^{15}$N~\cite{sao-np22,sao-prc22}. 

In the previous works, we studied various 
cases of elastic $\alpha$-$^{12}$C scattering 
at low energies~\cite{a-epja16,a-prc18,a-jkps18,a-prc22,a-prc23},
$E1$ transition of $^{12}$C($\alpha$,$\gamma$)$^{16}$O and an estimate of 
$S_{E1}$ factor of $^{12}$C($\alpha$,$\gamma$)$^{16}$O at $E_G$~\cite{a-prc19}, 
and $\beta$ delayed $\alpha$ emission from $^{16}$N~\cite{a-epja21}
up to the sub-leading order within the cluster EFT.  
The experimental data of each of the reactions 
are well reproduced by the fitted values of 
the parameters of reaction amplitudes, 
but a problem we observed 
in the previous work (see Fig.~6 in Ref.~\cite{a-prc22})
is that,
by using the fitted parameters to the precise phase shift data 
up to the $p$-$^{15}$N breakup energy, $E_\alpha = 6.62$~MeV
($E_\alpha$ is the $\alpha$ energy in the laboratory frame), 
reported by Tischhauser et al. (2009)~\cite{tetal-prc09},
a path of the inverse of dressed $^{16}$O propagator for $l=2$ 
cannot be uniquely determined in the low-energy region, 
where the $S_{E2}$ factor is extrapolated to $E_G$. 
In this work, we study this issue, by introducing conditions applied
to the effective range parameters in the low-energy region, 
employing two kinds of experimental data,
the phase shift of the elastic $\alpha$-$^{12}$C scattering
explicitly including resonant $2_4^+$ state of $^{16}$O and 
the $S_{E2}$ factor of $^{12}$C($\alpha$,$\gamma$)$^{16}$O 
below the energy of sharp resonant $2_2^+$ state of $^{16}$O. 

A known problem in the study of the elastic $\alpha$-$^{12}$C scattering 
for $l=2$ at low energy is that
the asymptotic normalization coefficient (ANC) of 
the subthreshold $2_1^+$ state of $^{16}$O calculated from 
the effective range parameters is significantly smaller than the values
deduced from other reactions, such as the $\alpha$ transfer reactions.
An estimate of the ANC of the subthreshold $2_1^+$ state of $^{16}$O, 
$|C_b|_2$, 
using the effective range parameters was reported by
K\"onig, Lee, and Hammer as 
$|C_b|_2 = (2.41\pm 0.38)\times 10^4$~fm$^{-1/2}$~\cite{klh-jpg13},
which is about a factor of five smaller than the value of 
$|C_b|_2 = (1.11\pm 0.11)\times 10^5$~fm$^{-1/2}$ 
deduced from the $\alpha$-transfer reactions, 
$^{12}$C($^{6}$Li,$d$)$^{16}$O and 
$^{12}$C($^7$Li,$t$)$^{16}$O~\cite{betal-prl99}.
While a large uncertainty of the ANC of the $2_1^+$ state deduced
from the elastic $\alpha$-$^{12}$C scattering
within a potential model, 
with values ranging from 2 to $18\times 10^4$~fm$^{-1/2}$, 
was reported by Sparenberg, Capel, and Baye~\cite{scb-jpcs11}. 
(The values of ANC of the $2_1^+$ state of $^{16}$O in the 
literature are summarized e.g. in Table 2 in Ref.~\cite{a-fb24}.)
As will be discussed in the following, 
the inverse of the dressed $^{16}$O propagator
for $l=2$ is represented in terms of the three effective range parameters,
$r_2$, $P_2$, $Q_2$, which approximately configure a cubic polynomial function. 
In Fig.~6 in Ref.~\cite{a-prc22}, 
three kinds of lines, 1) having a maximum point and a minimum point,
2) having a plateau, and 3) simply decreasing one, were obtained 
from the cubic function in the low energy region, 
where there are no data points
to determine which line is correct, even though those sets of fitted 
values of the effective range parameters evenly reproduce 
the accurate phase shift data well. 
Thus, we introduce the conditions to the effective range parameters, 
which make 
a value of the ANC of the $2_1^+$ state larger and 
the line of the inverse of the dressed $^{16}$O propagator for $l=2$
simply decreasing. 
Because there is no verification of the conditions,
we discuss its reliability by studying the effects of the conditions  
on the observable of the resonant $2_4^+$ state of 
$^{16}$O and 
the estimate of the $S_{E2}$ factor of $^{12}$C($\alpha$,$\gamma$)$^{16}$O
at $E_G$.

In this work, we first study the elastic $\alpha$-$^{12}$C scattering at low
energies including the resonant $2_4^+$ state of $^{16}$O in the cluster EFT.
A set of the experimental data of the phase shift 
up to $E_\alpha = 10$~MeV, 
reported by Bruno et al. (1975)~\cite{betal-nc75} is employed
along with the precise phase shift data reported 
by Tischhauser et al. (2009)~\cite{tetal-prc09}. 
The resonant $2_4^+$ state of $^{16}$O appears at 
$E_\alpha = \frac43 E_{R(24)} = 
7.9$~MeV, where $E_{R(24)}$ is the resonant energy of the $2_4^+$ state
of $^{16}$O, $E_{R(24)}=5.86$~MeV~\cite{twc-npa93}.
We introduce the conditions to restrict the parameter space of the effective 
range parameters in the low-energy region,
$E_\alpha = 0$ -- 2.6~MeV, and  
parameters of the $S$ matrix of the elastic $\alpha$-$^{12}$C scattering 
for $l=2$
are fitted to the two sets of the phase shift
data for two cases, with and without applying 
the conditions
to the effective range parameters.
For both cases, the fitted parameters reproduce the phase shift data
well, but we
find a large difference 
in the values of the ANC of the $2_1^+$ state of $^{16}$O;
we confirm that the ANC of the $2_1^+$ state of $^{16}$O cannot be determined
by the phase shift data of the elastic $\alpha$-$^{12}$C scattering for $l=2$.
We also find 
the noticeable differences in the values of parameters
for the resonant $2_4^+$ state of $^{16}$O. 
We compare the fitted values of the resonant energy and width 
of the $2_4^+$ state of $^{16}$O with those in the literature.

We then employ the experimental data of  
the $S_{E2}$ factor of $^{12}$C($\alpha$,$\gamma$)$^{16}$O. 
First, we study the energy dependence of the inverse 
of the dressed $^{16}$O propagator for $l=2$ in the low-energy region. 
We adjust the values of the effective range parameters for the large 
value of the ANC to reproduce the ANC of the $2_1^+$ state of $^{16}$O
deduced from the $\alpha$-transfer reactions. 
Then, using the fitted values of the effective range parameters
for the two cases, two additional parameters,
$y^{(0)}$ and $h_R^{(2)}$, of $E2$ transition amplitudes
of $^{12}$C($\alpha$,$\gamma$)$^{16}$O are fitted to the experimental data
of the $S_{E2}$ factor below the energy of the sharp resonant $2_2^+$ state 
of $^{16}$O.
We find the $\chi^2$ values as $\chi^2/N = 1.18$
and 1.55, for the cases with and without applying 
the conditions, respectively, 
where $N$ is the number of the data of the $S_{E2}$ factor.
Then, the $S_{E2}$ factor is extrapolated to $E_G=0.3$~MeV; we find
quite different values of the $S_{E2}$ factor at $E_G$.
We discuss the significance of introducing the conditions 
in the observables of the $2_4^+$ state of $^{16}$O
and the estimate of the $S_{E2}$ factor at $E_G$ and argue the necessity 
to adopt a value of the ANC of the $2_1^+$ state of $^{16}$O deduced from the 
$\alpha$-transfer reactions to reduce the uncertainty in
fixing the effective range parameters of the $2_1^+$ state of $^{16}$O.  

This paper is organized as follows. In Section 2, we review
the expression of the $S$ matrix of the elastic $\alpha$-$^{12}$C scattering
for $l=2$ in the cluster EFT. In Section 3, the numerical results
of this work are presented; in Section 3.1, the conditions applied to 
the effective range parameters are discussed, and in Section 3.2, the 
parameters of the $S$ matrix for $l=2$ 
are fitted to the experimental phase shift data,
and the fitted values of the resonant energy and width of the $2_4^+$ state
of $^{16}$O are compared with those in the literature. 
In Section 3.3, the energy dependence of 
the inverse of the dressed $^{16}$O propagator for $l=2$ on the conditions 
in the low energy region is studied.
Then, two additional parameters of the $E2$ transition amplitudes of 
$^{12}$C($\alpha$,$\gamma$)$^{16}$O are fitted to the experimental 
data of $S_{E2}$ factor and the $S_{E2}$ factor is extrapolated to $E_G$.
The numerical results of the $S_{E2}$ factor are presented and
discussed. 
Finally, in Section 4, the results and discussion of this
work are presented. 
In Appendix A, the expansion formulas of the digamma function
and the inverse of the dressed $^{16}$O propagator for $l=2$ are summarized,
and in Appendix B, the expression and derivation of the
$E2$ transition amplitudes of $^{12}$C($\alpha$,$\gamma$)$^{16}$O in 
the cluster EFT are briefly discussed. 

\vskip 2mm \noindent
{\bf 2. $S$ matrix of elastic $\alpha$-$^{12}$C scattering at low energies}

In this section, we review the expression of the $S$ matrices of 
the elastic $\alpha$-$^{12}$C scattering 
at low energies and its brief derivation
in the cluster EFT~\cite{a-prc23}. 
The $S$ matrices of the elastic $\alpha$-$^{12}$C scattering
for $l$th partial wave states are given in terms of
phase shifts, $\delta_l$, and elastic scattering amplitudes, $\tilde{A}_l$, as
\bea
S_l = e^{2i\delta_l}
= 1 + 2ip\tilde{A}_l\,.
\label{eq;S_l}
\eea
We now assume that the phase shifts can be decomposed as 
\bea
\delta_l = \delta_l^{(bs)} + \delta_l^{(rs1)}
+ \delta_l^{(rs2)}
+ \delta_l^{(rs3)}
\,,
\eea
where $\delta_l^{(bs)}$ is a phase shift generated from
a bound state, and $\delta_l^{(rsN)}$ with $N=1,2,3$
are those from the first, second, and third resonant states, and
each of those phase shifts may have
a relation to a corresponding scattering amplitude as
\bea
e^{2i\delta_l^{(ch)}} &=& 1 + 2ip\tilde{A}_l^{(ch)}\,,
\eea
where $ch(annel) = bs, rsN$,
and $\tilde{A}_l^{(bs)}$ and $\tilde{A}_l^{(rsN)}$ with $N=1,2,3$
are the amplitudes for the binding part and the first, second, 
and third resonant parts of the amplitudes,
which are derived from the effective Lagrangian in Ref.~\cite{a-prc23}.
The total amplitudes $\tilde{A}_l$ for the nuclear reaction part
in terms of the four amplitudes,
$\tilde{A}_l^{(bs)}$ and $\tilde{A}_l^{(rsN)}$
with $N=1,2,3$, read
\bea
\tilde{A}_l &=&
\tilde{A}_l^{(bs)}
+ e^{2i\delta_l^{(bs)}} \tilde{A}_l^{(rs1)}
+ e^{2i(\delta_l^{(bs)}+\delta_l^{(rs1)})} \tilde{A}_l^{(rs2)}
+ e^{2i(\delta_l^{(bs)}+\delta_l^{(rs1)}+\delta_l^{(rs2)})} \tilde{A}_l^{(rs3)}
\,.
\label{eq;tldAl}
\eea
We note that the total amplitudes
are not obtained as
the summation of the amplitudes,
but the additional phase factors appear in the front of them.

\begin{figure}[t]
\begin{center}
  \includegraphics[width=0.8\textwidth]{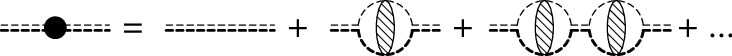}
\caption{
Diagrams for dressed $^{16}$O propagators.
A thick and thin double dashed line with or without a filled circle
represents the dressed or bare $^{16}$O propagator, respectively.
A thick (thin) dashed line represents a propagator of $^{12}$C ($\alpha$),
and a shaded blob in the loop diagrams the Coulomb green's function.
}
\label{fig:16O_propagator}
\label{fig;propagator}
\end{center}
\end{figure}
\begin{figure}[t]
\begin{center}
  \includegraphics[width=0.21\textwidth]{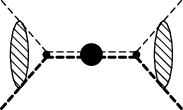}
\caption{
Diagram for elastic $\alpha$-$^{12}$C scattering amplitudes.
A shaded blob represents the initial or final Coulomb wave function,
and a thick and thin double-dashed line with a filled circle a
dressed $^{16}$O propagator. See the caption of Fig.~\ref{fig:16O_propagator}
as well.
}
\label{fig:Amplitude}
\end{center}
\end{figure}

The amplitudes are calculated by using the diagrams in
Figs.~\ref{fig:16O_propagator} and 
\ref{fig:Amplitude}~\cite{a-epja16,a-prc18,a-prc22}.
In the present study of the elastic $\alpha$-$^{12}$C scattering 
for $l=2$,
we include the subthreshold bound $2_1^+$ state and 
three resonant $2_2^+$, $2_3^+$, $2_4^+$ states of $^{16}$O.
For the bound state amplitude, $\tilde{A}_l^{(bs)}$ with $l=2$,
one has
\bea
\tilde{A}^{(bs)}_2 &=&
\frac{C_\eta^2 W_2(p)}{
K_2(p)
-2\kappa H_2(p)
}
\label{eq;Aer2}
\label{eq;A2_nr}
\,,
\eea
where $C_\eta^2 W_2(p)$ in the numerator of the amplitude
is calculated from the initial and final state Coulomb interactions
for $l=2$
in Fig.~\ref{fig:Amplitude}; $p$ is the magnitude of
relative momentum of the $\alpha$-$^{12}$C system in the center of mass
frame, $p=\sqrt{2\mu E}$, where $E$ is the energy of the $\alpha$-$^{12}$C
system, 
and
\bea
W_2(p) = 
\frac14
\left(\kappa^2 + 4 p^2\right) 
\left(\kappa^2 + p^2\right) \,, 
\ \ C_\eta^2 = \frac{2\pi\eta}{\exp(2\pi\eta)-1}\,,
\eea
where $\eta$ is the Sommerfeld parameter, $\eta=\kappa/p$:
$\kappa$ is the inverse of the Bohr radius,
$\kappa = Z_\alpha Z_{12C} \alpha_E\mu$,
where $Z_A$ is the number of protons
in a nucleus, and $\alpha_E$ is the fine structure constant.
One has $\kappa=245$~MeV, which is regarded as a large scale of the theory
because of $\kappa > \Lambda_H$.
The function $-2\kappa H_2(p)$ in the denominator of the amplitude
is the Coulomb self-energy term which is calculated
from the loop diagram in Fig.~\ref{fig:16O_propagator}, and one has
\bea
H_2(p) &=& W_2(p) H(\eta)\, ,
\ \ \
H(\eta) = \psi(i\eta) + \frac{1}{2i\eta} -\ln(i\eta)\,,
\label{eq;H2}
\eea
where $\psi(z)$ is the digamma function.
As discussed in Ref.~\cite{a-prc18}, large and significant contributions
to the series of effective range expansions,
compared to the terms calculated using a phase shift datum at the lowest energy
of the data, $E_\alpha=2.6$~MeV~\cite{tetal-prc09},
appear from the Coulomb self-energy term, $-2\kappa H_l(p)$
with $l=0,1,2$. 
In addition, for $l=2$, one can find the appearance of the large
terms by expanding the self-energy term, $2\kappa H_2(p)$, in terms
of $1/\eta^2=(p/\kappa)^2$ in $p\to 0$ limit.
Expressions of the function $H_2(p)$ expanded in powers of $(p/\kappa)^2$ 
are presented in Appendix A. Thus, one has
\bea
2\kappa ReH_2(p) = \frac{1}{24}\kappa^3 p^2 
+ \frac{17}{80}\kappa p^4 
+ \frac{757}{4032\kappa}p^6
+ \frac{289}{10080\kappa^3}p^8 
+ \frac{491}{22176\kappa^5}p^{10}
+ \cdots\,,
\label{eq;2kappaReH2}
\eea
where one may notice that the large terms proportional to $\kappa^3$
and $\kappa$ appear in the first and second terms on the right-hand-side
of the equation. Those terms are regarded as the terms which do not obey the 
counting rules and need to be subtracted by counter 
terms~\cite{gss-npb88,af-prd07}.

Nuclear interaction is represented
in terms of the effective range parameters
in the function $K_2(p)$ in
the denominator of the amplitude in Eq.~(\ref{eq;Aer2}).
We introduce two terms proportional to $p^2$ and $p^4$ as leading
order contributions, to subtract the two large contributions from the 
self-energy term mentioned above,
and a term proportional to $p^6$ as a sub-leading 
one; the effective range terms up to
$p^6$ order are included for $l=2$, 
and we have
\bea
K_2(p) &=&
-\frac{1}{a_2}
+\frac12 r_2p^2
-\frac14 P_2 p^4
+Q_2 p^6
\,,
\eea
where $a_2$, $r_2$, $P_2$, $Q_2$ are the effective range parameters
for $l=2$.

We fix a parameter among the four effective range parameters,
$a_2$, $r_2$, $P_2$, $Q_2$, by using a condition that
the inverse of the scattering amplitude $\tilde{A}_2^{(bs)}$ vanishes at
the binding energy of the $2_1^+$ state of $^{16}$O.
Thus, the denominator of the scattering amplitude,
\bea
D_2(p) = K_2(p) -2\kappa H_2(p) \,,
\label{eq;binding_pole}
\label{eq;D2}
\eea
vanishes at $p=i\gamma_{2}$ where $\gamma_{2}$ are the binding momentum
of the $2_1^+$ state of $^{16}$O;
$\gamma_{2} = \sqrt{2\mu B_{2}} = 37.0$~MeV. 
We fix the scattering length $a_2$ by using the condition and rewrite 
the expression of the function $K_2(p)$ as
\bea
K_2(p) &=& \frac12r_2(\gamma_2^2 + p^2) + \frac14P_2(\gamma_2^4-p^4)
+Q_2(\gamma_2^6+p^6)
+ 2\kappa H_2(i\gamma_2)
\,.
\eea

At the binding energy, one has the wave function normalization factor
$\sqrt{Z_{2}}$ for the bound $2_1^+$ state of $^{16}$O in the dressed $^{16}$O
propagator for $l=2$ as
\bea
\frac{1}{D_2(p)} = \frac{Z_{2}}{E+B_{2}} + \cdots\,,
\eea
where the dots denote the finite terms at $E=-B_{2}$, and
one has
\bea
\sqrt{Z_{2}} = \left(
\left|
\frac{dD_2(p)}{dE}
\right|_{E=-B_{2}}
\right)^{-1/2}=
\left(
2\mu
\left|
\frac{dD_2(p)}{dp^2}
\right|_{p^2=-\gamma_{2}^2}
\right)^{-1/2}\,.
\label{eq;sqrtZl}
\eea
The wave function normalization factor $\sqrt{Z_{2}}$ is multiplied to a
reaction amplitude when the bound state appears in the initial or
final state of a reaction.

The ANCs $|C_b|_{l}$ for the bound states of $^{16}$O are calculated
by using the formula of Iwinski, Rosenberg, and Spruch~\cite{irs-prc84}
\bea
|C_b|_{l} = \frac{\gamma_{l}^l}{l!} \Gamma(l+1+\kappa/\gamma_{l})
\left( \left| \frac{dD_l(p)}{dp^2}\right|_{p^2 = - \gamma_{l}^2}
\right)^{-1/2} \ \,, 
\label{eq;Cb}
\eea
where $\Gamma(x)$ is the gamma function, and one may notice that the ANCs
are proportional to the wave function normalization factor $\sqrt{Z_{l}}$
comparing Eqs.~(\ref{eq;sqrtZl}) and (\ref{eq;Cb}). 
The ANC of the $2_1^+$ state 
of $^{16}$O, $|C_b|_2$, can be calculated by using the fitted values of
the effective range parameters, $r_2$, $P_2$, $Q_2$. 

The amplitudes for the resonant $2_2^+$, $2_3^+$, $2_4^+$ states may be 
obtained in the Breit-Wigner-like expression as
\bea
\tilde{A}_2^{(rsN)} &=&
-
\frac{1}{p}
\frac{\frac12\Gamma_{(2i)}(E) }{E-E_{R(2i)}
+ R_{(2i)}(E) + i\frac12\Gamma_{(2i)}(E)}\,,
\label{eq;A2_rsN}
\eea
with
\bea
\Gamma_{(2i)}(E) &=& \Gamma_{R(2i)}
\frac{pC_\eta^2W_2(p)}
     {p_rC_{\eta_r}^2W_2(p_r)}\,,
\\
R_{(2i)}(E) &=& a_{(2i)}(E-E_{R(2i)})^2 + b_{(2i)}(E-E_{R(2i)})^3
\,,
\label{eq;R}
\eea
where $E_{R(2i)}$ and $\Gamma_{R(2i)}$ are the energy and width of
the resonant $2_{i}^+$ states (where $i=N+1$ with $N=1,2,3$), 
and $p_r$ and $\eta_r=\kappa/p_r$ are
the momenta and Sommerfeld factors at the resonant energies:
we suppressed the $i$ indices for them.
The functions $R_{(2i)}(E)$ have
the second and third order corrections expanded around $E=E_{R(2i)}$,
where the coefficients, $a_{(2i)}$ and $b_{(2i)}$, are fitted
to the shapes of resonant peaks.

Using the relations for the amplitudes
in Eqs.~(\ref{eq;A2_nr}) and (\ref{eq;A2_rsN}),
the $S$ matrix for $l=2$ in Eq.~(\ref{eq;S_l}) is obtained
in a simple and transparent expression as
\bea
e^{2i\delta_2} &=&
\frac{K_2(p) - 2\kappa Re H_2(p) + ipC_\eta^2W_2(p)}
     {K_2(p) - 2\kappa Re H_2(p) - ipC_\eta^2W_2(p)}
\nnb \\ && \times
\prod_{i=2}^4
\frac{E - E_{R(2i)} + R_{(2i)}(E) - i\frac12\Gamma_{(2i)}(E)}
     {E - E_{R(2i)} + R_{(2i)}(E) + i\frac12\Gamma_{(2i)}(E)}
\,,
\label{eq;exp2idel_l}
\eea
where we represented the part of the subthreshold state 
as a function of momentum, $p$, and the parts of the resonant states
as functions of energy, $E$; they are related by the non-relativistic 
equation, $E=p^2/(2\mu)$. 

\vskip 2mm \noindent
{\bf 3. Numerical results}

In this section, we first introduce the conditions to apply to the effective
range parameters when fitting them to the phase shift data. 
We then employ two kinds of experimental data,
the phase shift of the elastic $\alpha$-$^{12}$C scattering 
for $l=2$ up to $E_\alpha= 10$~MeV and 
the $S_{E2}$ factor of $^{12}$C($\alpha$,$\gamma$)$^{16}$O up to $E=2.5$~MeV. 
Employing the phase shift data, we fit the parameters of the $S$ matrix 
of the elastic $\alpha$-$^{12}$C scattering for $l=2$ with and without applying 
the conditions, and compare the fitted values of resonant energy and width 
of the $2_4^+$ state of $^{16}$O with those in the literature.
We then study the energy dependence of the inverse of 
the dressed $^{16}$O propagator for $l=2$ in
the low-energy region by using the fitted values of the effective range
parameters. Then, employing the experimental data of the $S_{E2}$ factor,
we fit additional parameters of the $E2$ transition amplitudes of 
$^{12}$C($\alpha$,$\gamma$)$^{16}$O, and the $S_{E2}$ factor is extrapolated
to $E_G$. 

\vskip 1mm \noindent
{\bf 3.1 Conditions applied to the effective range parameters}

The inverse of the propagator,
$D_2(p)$, is approximately represented as a cubic equation in powers of $p^2$,
whose coefficients are given by the effective range parameters
$r_2$, $P_2$, $Q_2$. In general,
it can have a minimum point and a maximum point, 
a flat plateau,
or a simply decreasing one in the low-energy region,
as mentioned above. 
To make it a simple decreasing function,
which results in a large value of the ANC of the $2_1^+$ state of $^{16}$O, 
we introduce the conditions
when fitting the effective range parameters, 
$r_2$, $P_2$, $Q_2$, in the low energy region, $0\le E_\alpha \le 2.6$~MeV.

We first expand the function $H(\eta)$ in terms of $1/\eta$ 
in the asymptotic limit, $\eta \to \infty$;
the formulas for the expansion of the digamma function $\psi(z)$
are summarized in Appendix A. 
Thus, 
the real part of the inverse of the propagator, 
$ReD_2(p)$, in Eq.~(\ref{eq;D2}) expanded around the binding energy, 
$E=-B_2$, i.e., $p^2 = -\gamma_2^2$, is obtained as
\bea
ReD_2(p) \simeq \sum_{n=1}^5C_n (\gamma_2^2 + p^2)^n\,,
\label{eq;ReD2}
\eea
with
\bea
C_1 &=& \frac12\left(
r_2 - \frac{1}{12}\kappa^3
\right) + \frac12\left(
P_2 + \frac{17}{20}\kappa
\right)\gamma_2^2
+ 3\left(
Q_2 - \frac{757}{4032\kappa}
\right)\gamma_2^4
\nnb \\ && 
+ \frac{289}{2520\kappa^3}\gamma_2^6 
- \frac{2455}{22176\kappa^5}\gamma_2^8 
+ \cdots\,,
\label{eq;C1}
\\
C_2 &=& -\frac14\left(
P_2 + \frac{17}{20}\kappa
\right) - 3\left(
Q_2 - \frac{757}{4032\kappa}
\right)\gamma_2^2
- \frac{289}{1680\kappa^3}\gamma_2^4
+ \frac{2455}{11088\kappa^5}\gamma_2^6
- \cdots\,,
\label{eq;C2}
\\
C_3 &=& Q_2 - \frac{757}{4032\kappa}
+ \frac{289}{2520\kappa^3}\gamma_2^2
- \frac{2455}{11088\kappa^5}\gamma_2^4
+ \cdots\,,
\label{eq;C3}
\\
C_4 &=& - \frac{289}{10080\kappa^3} + \frac{2455}{22176\kappa^5}\gamma_2^2
- \cdots\,,
\label{eq;C4}
\\
C_5 &=& - \frac{491}{22176\kappa^5} + \cdots\,,
\label{eq;C5}
\eea
and the conditions; $C_n<0$ for $n=1, 2, 3$ are introduced, 
which make $ReD_2(p)$ simply decrease in the low energy region. 
(One may notice that $C_4$, $C_5<0$ above.)
These conditions lead to restrictions to
the effective range parameters as
\bea
Q_2 &<& \frac{757}{4032\kappa}
- \frac{289}{2520\kappa^3}\gamma_2^2
+ \frac{2455}{11088\kappa^5}\gamma_2^4 
+\cdots\,,
\label{eq;Q2_cond}
\\
P_2 &>& - \frac{17}{20}\kappa 
- 12\left(Q_2 -\frac{757}{4032\kappa}
\right)\gamma_2^2 
- \frac{289}{420\kappa^3}\gamma_2^4 
+ \frac{2455}{2772\kappa^5}\gamma_2^6 
+ \cdots \,,
\label{eq;P2_cond}
\\
r_2 &<& \frac{1}{12}\kappa^3 
-\left(P_2 + \frac{17}{20}\kappa
\right)\gamma_2^2
-6\left(Q_2 - \frac{757}{4032\kappa}
\right)\gamma_2^4
- \frac{289}{1260\kappa^3}\gamma_2^6
+ \frac{2455}{11088\kappa^5}\gamma_2^8
+ \cdots\,,
\label{eq;r2_cond}
\eea
where the terms are expanded in powers of $(\gamma_2/\kappa)^2 = 0.023$
[$<(Q/\Lambda_H)^2 = 0.0625$];
the truncation of higher-order terms would be a good approximation. 
From those conditions, 
one has the minimum or maximum values of the effective range parameters
as
\bea
r_{2,max}=0.159026\, \textrm{fm}^{-3}\,, \ \ \ 
P_{2,mim}=-1.05390\, \textrm{fm}^{-1}\,, \ \ \ 
Q_{2,max}=0.149343\, \textrm{fm}\,.  
\label{eq;restrictions_on_parameters}
\eea
We note that the wave function normalization factor 
$Z_2$ in Eq.~(\ref{eq;sqrtZl}) is obtained by $C_1$ 
in Eq.~(\ref{eq;C1}), $Z_2^{-1} = 2\mu C_1$, and the ANC of 
the $2_1^+$ state of $^{16}$O is presented as
\bea
|C_b|_2 = \frac12\gamma_2^2\,\Gamma(3+\kappa/\gamma_2)
\frac{1}{\sqrt{C_1}}\,.
\label{eq;ANCnr2}
\eea
Thus, if one adopts the ANC of $2_1^+$ state of $^{16}$O,
$|C_b|_2$, as an input, then one can fix one of the three effective
range parameters, $r_2$, $P_2$, $Q_2$, in $C_1$ by this equation. 

\vskip 1mm \noindent
{\bf 3.2 Fitting the effective range parameters and 
the $2_4^+$ state of $^{16}$O}

In the previous work~\cite{a-prc22}, we employed the precise phase shift data
up to $E_\alpha = 6.62$~MeV, reported
by Tischhauser et al. (2009)~\cite{tetal-prc09}, 
to fit the parameters including the 
resonant $2_2^+$ and $2_3^+$ states of $^{16}$O. (They appear at 
$E_\alpha(2_2^+) = 3.58$~MeV and $E_\alpha(2_3^+)=5.81$~MeV.)
We obtained six sets of the values of effective range parameters, 
$r_2$, $P_2$, $Q_2$, fitted well the precise phase shift data for $l=2$
(see TABLEs I and II in Ref.~\cite{a-prc23}), but those values make 
the different values of the ANC of the $2_1^+$ state of $^{16}$O and the 
different
paths of the real part of the inverse of the dressed $^{16}$O propagator 
for $l=2$,
$ReD_2(p)$, in the low energy region where the $S_{E2}$ factor
is extrapolated to $E_G$ (see Fig.~6 in Ref.~\cite{a-prc23}). 
In the present work, we employ and include a set of the phase shift data for
$l=2$ up to $E_\alpha = 10$~MeV reported
by Bruno et al. (1975)~\cite{betal-nc75}, 
to refit the parameters explicitly including the resonant $2_4^+$ state
of $^{16}$O in the $S$ matrix of the elastic $\alpha$-$^{12}$C scattering 
for $l=2$. 
There are 13 parameters, $\theta =\{r_2, P_2, Q_2, 
E_{R(22)}, \Gamma_{R(22)}, E_{R(23)}, \Gamma_{R(23)}, a_{(23)}, b_{(23)},
E_{R(24)}, \Gamma_{R(24)}, a_{(24)}, b_{(24)}
\}$, which are fitted to the two sets of the phase shift data,
introducing the conditions to the effective range parameters,
by means of the $\chi^2$ fit using an MCMC ensemble sampler~\cite{mcmc}. 

\begin{table}
\begin{center}
\begin{tabular}{ l | l l l }
\hline 
	& Prev. work & This work & This work \cr 
	& w/o cond. & w/o cond. & w cond. \cr \hline 
	$r_2$~(fm$^{-3}$) & 0.149(4) & 0.150(6) & 0.1586(3) \cr 
	$P_2$~(fm$^{-1}$) & -1.19(5) & -1.18(8) & -1.047(2) \cr
	$Q_2$~(fm) & 0.081(16) & 0.084(3) & 0.138(2) \cr
	$E_{R(22)}$~(MeV) & 2.68308(5) & 2.68308(1) &  2.68308(1) \cr
	$\Gamma_{R(22)}$~(keV) & 0.75(2) & 0.76(1) & 0.76(1) \cr
	$E_{R(23)}$~(MeV) & 4.3545(2) & 4.3533(3) & 4.3537(1) \cr
	$\Gamma_{R(23)}$~(keV) & 74.61(3) & 74.5(1) & 74.5(1) \cr
	$a_{(23)}$~(MeV$^{-1}$) & 0.46(12) & 0.6(2) & 1.1(1) \cr
	$b_{(23)}$~(MeV$^{-2}$) & 0.47(9) & 0.5(2) & 0.6(1) \cr
	$E_{R(24)}$ (MeV) & 5.858$^*$ & 5.92(2) & 5.90(2) \cr
	$\Gamma_{R(24)}$ (keV) & 150$^*$ & 300$^{+60}_{-40}$ & 235(20) \cr
	$a_{(24)}$~(MeV$^{-1}$) & -- & 0.3(4) & 0.6(1) \cr
	$b_{(24)}$~(MeV$^{-2}$) & -- & 0.96$^{+0.79}_{-0.50}$ & 0.3(1) \cr
	\hline
	$|C_b|_2$ (fm$^{-1/2}$) & 3.1(6)$\times 10^4$ & 3.24$\times 10^4$ & 
	22.8$\times 10^4$ \cr
	$\chi^2/N$ ($N$) & 0.66 (245) & 3.02 (271) & 3.04 (271) \cr \hline
\end{tabular}
\caption{
	Fitted values of the parameters of the $S$ matrix
	of the elastic $\alpha$-$^{12}$C scattering for $l=2$
	to the two sets of the phase shift data~\cite{tetal-prc09,betal-nc75}.
	In the second column, those from the column (I) in Table II in the
	previous work~\cite{a-prc22}, in the third column, those of this
	work without applying the conditions,
	and in the fourth column, those of this work 
	applying the conditions to the effective range parameters 
	in Eqs.~(\ref{eq;Q2_cond}), (\ref{eq;P2_cond}), (\ref{eq;r2_cond}) are
	displayed. 
    In the second row from the bottom, values of the ANC of the $2_1^+$
	state of $^{16}$O, which are calculated with the values of $r_2$, $P_2$,
	$Q_2$, and in the last row, values of $\chi^2/N$ ($N$) 
	($N$ are the numbers of data), are displayed. 
	In the previous work, 
	$E_{R(24)}$ and $\Gamma_{R(24)}$ were included as fixed values 
	(marked by $*$) 
	by using the experimental data~\cite{twc-npa93}, 
	and the parameters $a_{(24)}$ and $b_{(24)}$ were not included. 
}
\label{table;parameters}
\end{center}
\end{table}
In Table \ref{table;parameters}, 
values of the parameters fitted to the phase shift data are displayed; 
in the second column those 
in the previous work (the column (I) in TABLE II)~\cite{a-prc22}, 
in the third column those of this work without applying the conditions,
and in the fourth column, those of this work applying the conditions 
in Eqs.~(\ref{eq;Q2_cond}), (\ref{eq;P2_cond}), (\ref{eq;r2_cond}) to 
the effective range parameters, are presented. 
In the previous work, we employed the 
experimental data reported by Tischhauser et al. (2009)~\cite{tetal-prc09} only
(the number of data is $N=245$);
we included the $2_4^+$ state as a background from high energy where the 
resonant energy and width are fixed by using the experimental 
data~\cite{twc-npa93} and the 
parameters $a_{(24)}$ and $b_{(24)}$ were not included. One can see that 
the values in the second and third columns are in good agreement except
for those of $\Gamma_{R(24)}$ and $\chi^2/N$. 
We discuss the values of $\Gamma_{R(24)}$ later, and the larger values of
$\chi^2/N$ are due to the inclusion of the phase shift data reported 
by Bruno et al. (1975)~\cite{betal-nc75} 
(the number of data is $N=26$). 
We find that 
the conditions applied to the effective range parameters change the 
values of $r_2$, $P_2$, $Q_2$ significantly in the third and fourth 
columns. 
One may notice that the values of the effective range parameters
in the second and third columns do not satisfy the bounds due to 
the conditions in Eq.~(\ref{eq;restrictions_on_parameters}),
and those in the last column satisfy them. 
The values of ANC, $|C_b|_2$ are altered largely;
we obtain $|C_b|_2 = 3.24\times 10^4$~fm$^{-1/2}$ when not applying the 
conditions, which is about a factor of 1.6 larger than that reported by
K\"onig, Lee, and Hammer. This may be due to the inclusion of $2_4^+$
state of $^{16}$O (see Ref.~\cite{a-prc22} as well),
and $|C_b|_2 = 22.8\times 10^4$~fm$^{-1/2}$ when applying the conditions, 
which is about a factor of two larger than 
those deduced from the $\alpha$-transfer reactions. 
While this range of the values of the ANC, $|C_b|_2 = (3.2 -22.8)\times 
10^4$~fm$^{-1/2}$ agrees with that reported by Sparenberg, Capel, and Baye
in their study employing a potential model, 
$(2-18)\times 10^4$~fm$^{-1/2}$~\cite{scb-jpcs11}. 
In addition, the values of $\chi^2/N$ are similar
in the third and fourth columns; 
we confirm that the ANC of the $2_1^+$ state of $^{16}$O cannot be
determined by the phase shift data of the elastic $\alpha$-$^{12}$C 
scattering for $l=2$. 
One can also find that the values of the shape parameter, $a_{(23)}$, 
of the $2_3^+$ state and the width and shape parameters,
$\Gamma_{R(24)}$, $a_{(24)}$, and $b_{(24)}$ of the $2_4^+$ state are 
altered between the third and fourth columns in the table. 

\begin{figure}[t]
\begin{center}
\resizebox{0.8\textwidth}{!}{
 \includegraphics{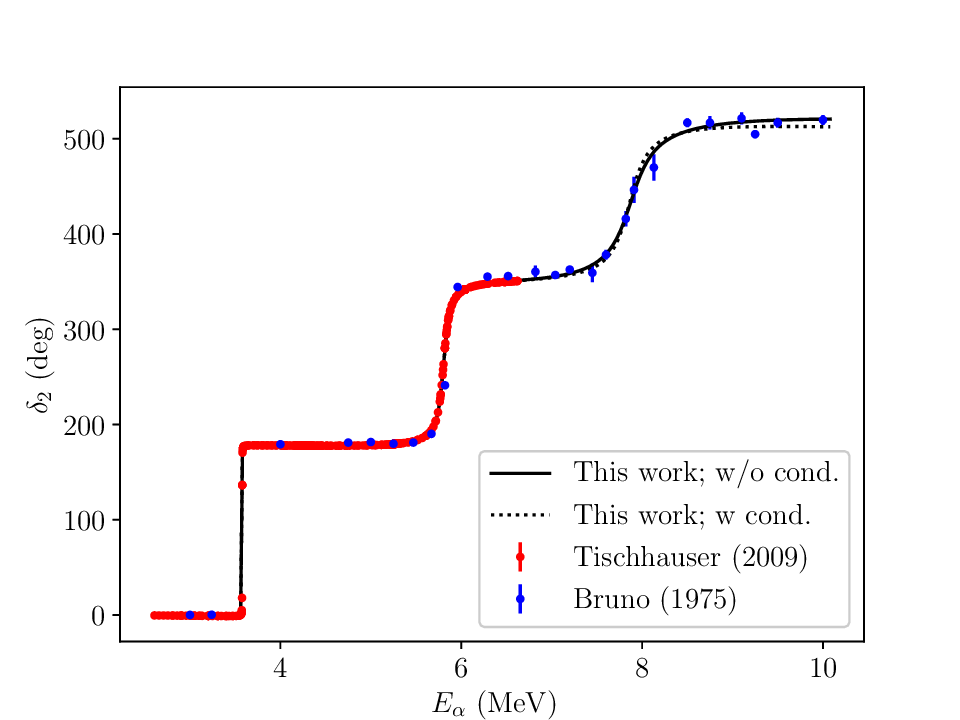}
}
\caption{
	Phase shift of the elastic $\alpha$-$^{12}$C scattering for $l=2$ 
	plotted as a function of the $\alpha$
	energy $E_\alpha$ in the laboratory frame. A line is plotted by using
	the values of the parameters in the third column 
	in Table~\ref{table;parameters} and a dotted line by using those in 
	the fourth column in the same table. The experimental data reported
	by Tischhauser et al. (2009)~\cite{tetal-prc09} 
	and Bruno et al. (1975)~\cite{betal-nc75} are displayed 
	in the figure as well. 
}
\label{fig;del2}
\end{center}
\end{figure}
In Fig.~\ref{fig;del2}, 
the phase shift of the elastic $\alpha$-$^{12}$C scattering
for $l=2$ are plotted as a function of the $\alpha$ energy $E_\alpha$. 
A solid line is plotted using the values of the parameters in the third
column in Table~\ref{table;parameters}, 
and a dotted line is drawn using those in the fourth column of the same table. 
The experimental data reported by
Tischhauser et al. (2009)~\cite{tetal-prc09} (the accurate data up to
the $p$-$^{15}$N breakup energy, $E_\alpha=6.62$~MeV) 
and Bruno et al. (1975)~\cite{betal-nc75} 
(the data covering the high-energy region
for the resonant $2_4^+$ state of $^{16}$O up to $E_\alpha = 10$~MeV) are
also displayed in the figure.
One can see that both lines reproduce the experimental data well. 

\begin{figure}
\begin{center}
  \includegraphics[width=13cm]{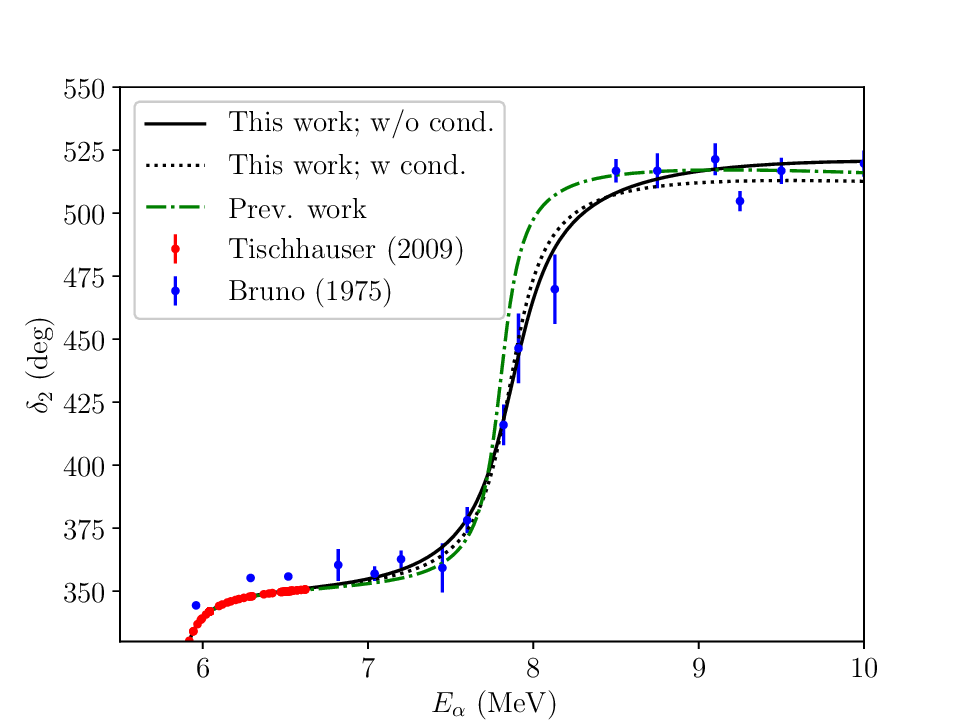}
\caption{
	The same phase shift displayed in Fig.~\ref{fig;del2} 
	in the energy region of the resonant $2_4^+$ state plotted 
	as a function of the $\alpha$ energy $E_\alpha$. 
	A dashed-dotted line is also plotted using the parameters obtained in the previous work (those in the second column in Table \ref{table;parameters}). 
	See the caption in Fig.~\ref{fig;del2} as well. 
}
\label{fig;del2_24p}       
\end{center}
\end{figure}
In Fig.~\ref{fig;del2_24p}, 
the same lines and data 
shown in Fig.~\ref{fig;del2} are displayed
in the energy region for the resonant $2_4^+$ state of $^{16}$O.
A dashed-dotted line using the parameters obtained in the previous work
(those in the second column in Table~\ref{table;parameters}) is also
plotted in the figure. 
One can see that the lines fitted to the data in this work become better 
than that in the previous work. 
The two lines in this work are distinguishable, 
but the data have a significant size of the error bars; 
it may not be easy to determine which line is better than
the other. 
As discussed above, this difference can also be seen in
the different values of the
parameters of the $2_4^+$ state of $^{16}$O 
in the third and fourth columns of Table~\ref{table;parameters}. 

\begin{table}
\begin{center}
\begin{tabular}{ l | l l l l l }
\hline 
	& Bruno & TWC & deBoer & This work &  \cr
	& (1975) & (1993) & (2012) & w/o cond. & w cond.  
	\cr \hline 
	$E_{R(24)}$~(MeV) & 5.83(3) & 5.858(10) & 5.805(2) & 5.92(2) & 
	5.90(2)  \cr
	$\Gamma_{R(24)}$~(keV) & 520(200) & 150(10) & 349(3) & 
	300$^{+60}_{-40}$ & 235(20) \cr \hline
\end{tabular}
\caption{
	Resonant energy and width of the $2_4^+$ state of $^{16}$O.
The values in the second, third, and fourth columns are from the literature;
Bruno et al. (1975)~\cite{betal-nc75}, 
the compilation edited
by Tilley, Weller, and Cheves (TWC) (1993)~\cite{twc-npa93}, 
and deBoer et al. (2012)~\cite{detal-prc12}, respectively. 
	Those in the fifth and sixth columns are the fitted values of 
	this work without and with the conditions applied to the effective
	range parameters. 
}
\label{table;E24rnG24r}
\end{center}
\end{table}
In Table~\ref{table;E24rnG24r}, we summarize the values of resonant energy 
and width of the $2_4^+$ state of $^{16}$O, $E_{R(24)}$ and $\Gamma_{R(24)}$,
in the literature and our results 
presented in Table~\ref{table;parameters}. 
We have larger values of the resonant energy, $E_{R(24)}$, by two sigma
deviation from the value of Bruno et al. (1975)~\cite{betal-nc75}.
One can see that the values of $\Gamma_{R(24)}$ in the literature are
still scattered and the uncertainties of those values are significant;
those values are in good agreement within the error bars except for that
of the compilation edited by Tilley, Weller, 
and Cheves (1993)~\cite{twc-npa93},
$\Gamma_{R(24)}=150(10)$~keV, which is significantly smaller 
than the others. 
To improve the situation,
it may be helpful to have a more precise data set of the phase shift 
in the energy region for the resonant $2_4^+$ state of $^{16}$O. 
We note that because
two channels, $\alpha$-$^{12}$C$^*$($2_1^+$) and $p$-$^{15}$N states, 
open in this energy region, 
the inelastic channels of the scattering start contributing.
Thus, it is necessary to improve the treatment in the theory as well. 

\vskip 1mm \noindent
{\bf 3.3 Dressed $^{16}$O propagator and
and the estimate of $S_{E2}$ factor at $E_G$}

\begin{figure}
\begin{center}
  \includegraphics[width=13cm]{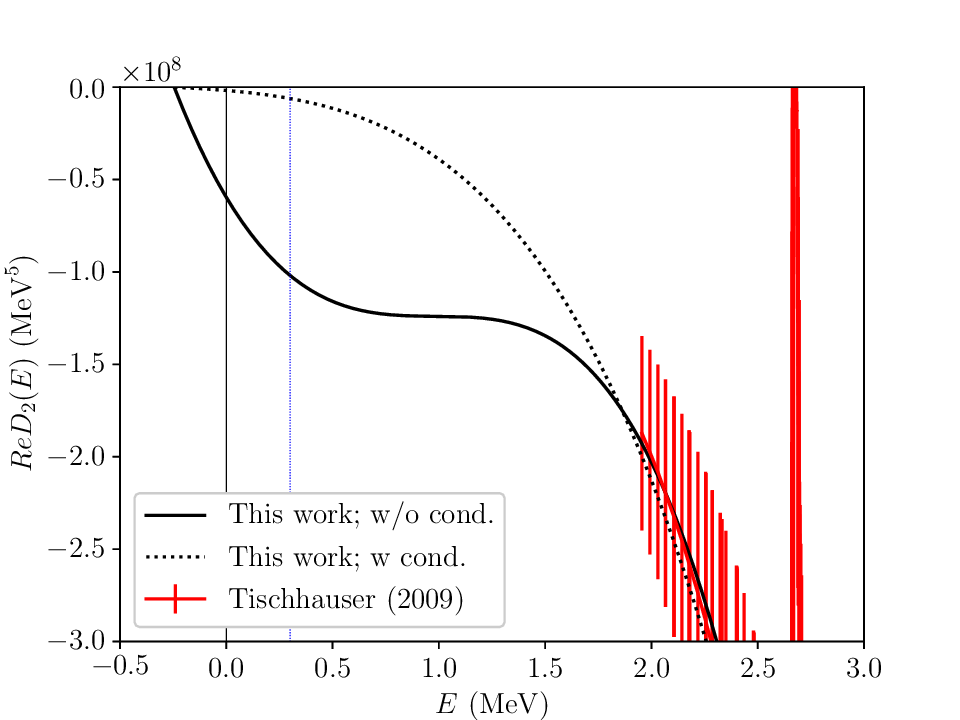}
	\caption{Real part of the inverse of the propagator, $ReD_2(E)$
	[$=ReD_2(p)$], 
	plotted as a function of the energy $E$ of the $\alpha$-$^{12}$C 
	system in the center-of-mass frame. A solid line is plotted 
	using the values of the effective range parameters, 
	$r_2$, $P_2$, $Q_2$, in the third column 
	in Table~\ref{table;parameters} and
	a dotted line by those in the fourth column in the same table.
	The phase shift data reported 
	by Tischhauser et al. (2009)~\cite{tetal-prc09} are converted 
	and displayed in the figure as well. A vertical blue line is drawn at 
	$E_G = 0.3$~MeV. 
}
\label{fig;ReD2}       
\end{center}
\end{figure}
We are now in the position to study the effect of the conditions
applied to the effective range parameters on the inverse of the 
propagator, $D_2(p)$, 
and the calculation of the $S_{E2}$ factor of the $E2$ transition 
of $^{12}$C($\alpha$,$\gamma$)$^{16}$O. 
In Fig.~\ref{fig;ReD2},
we plot the real part of $D_2(E)$ [$=D_2(p)$] 
as a function of the energy $E$ 
of the $\alpha$-$^{12}$C system in the center-of-mass frame
at the low-energy region. 
A solid line is 
calculated by using the values of $r_2$, $P_2$, $Q_2$ in the third column 
of Table~\ref{table;parameters} and a dotted line by using those in the 
fourth column of the same table. The experimental data of the phase
shift reported by Tischhauser et al. (2009)~\cite{tetal-prc09}
are converted to $ReD_2(E)$ using a relation,
\bea
ReD_2(p) = p W_2(p)C_\eta^2\cot \delta_2\,,
\eea
and plotted in the figure as well. 
One can see that the paths of the two lines are quite different
because of the conditions
applied (or not applied) to the effective range parameters. 
The solid line has
a plateau in the low energy region, $0<E<1.95$~MeV, and the dotted line is 
smoothly decreasing, while both lines reproduce the experimental data 
equally well.
In addition, at the top of the figure, both the lines start at the 
same point, i.e., at the binding energy of the $2_1^+$ state of $^{16}$O, 
$E=-B_2$, where $D_2(-B_2)=0$. 
One may notice that 
the gradients of the lines at this point are also quite different;
they are related to the values of 
the ANC of the $2_1^+$ state of $^{16}$O, $|C_b|_2$, in Eq.~(\ref{eq;Cb}).
Because the square of the root of the gradient appears 
in the denominator of the 
formula of $|C_b|_2$, a large angle associated with the horizontal line
at this point 
leads to a small value of the ANC, 
and a small angle leads to a large value of the ANC. 
Thus, we obtained quite different values, 
the small and large values of the ANC
in Table~\ref{table;parameters}. 
The two lines go through 
the different paths between the point at $E=-B_2$ and the datum 
of phase shift whose lowest energy is $E=\frac34 E_\alpha = 1.95$~MeV. 
Because the inverse of the propagator $D_2(p)$ appears 
in the denominator of the $E2$ transition amplitudes 
of $^{12}$C($\alpha$,$\gamma$)$^{16}$O, 
the energy dependence of $D_2(E)$ [$=D_2(p)$] in the low energy region 
is crucial
when extrapolating the $S_{E2}$ factor to
$E_G=0.3$~MeV. 

By employing the two sets of the fitted values of 
the effective range parameters
in Table~\ref{table;parameters}, 
we fit additional parameters
in the $E2$ transition amplitudes of 
$^{12}$C($\alpha$,$\gamma$)$^{16}$O to the experimental data 
of the $S_{E2}$ factor of $^{12}$C($\alpha$,$\gamma$)$^{16}$O where
we have adjusted the values of the effective range parameters
for the large ANC 
in the fourth column of Table \ref{table;parameters}
to reproduce the ANC of the $2_1^+$ state of $^{16}$O 
deduced from the $\alpha$-transfer reactions, 
$|C_b|_2 = 10\times 10^4$~fm$^{-1/2}$, with Eq.~(\ref{eq;ANCnr2}); 
we have
\bea
r_2 = 0.157453~\textrm{fm}^{-3}\,, \ \ 
P_2 = -1.0481~\textrm{fm}^{-1}\,, \ \ 
Q_2 = 0.1403~\textrm{fm}\,.
\label{eq;r2P2Q2}
\eea
The expression of the $E2$ transition amplitudes 
and its brief derivation are presented 
in Appendix B; we have two additional parameters, 
$h_R^{(2)}$ and $y^{(0)}$, in the amplitudes. 
The experimental data of the $S_{E2}$ factor 
below the resonant energy of $2_2^+$ of $^{16}$O, up to $E=2.5$~MeV, 
reported by Ouellet et al. (1996)~\cite{oetal-prc96}, 
Roters et al. (1999)~\cite{retal-epja99}, 
Kunz et al. (2001)~\cite{ketal-prl01},
Fey (2004)~\cite{f-04}, 
Makii et al. (2009)~\cite{metal-prc09}, 
and Plag et al. (2012)~\cite{petal-prc12}, 
are employed for fit.

\begin{table}
\begin{center}
\begin{tabular}{ l | l l }
\hline 
	& This work (w/o cond.) & This work (w cond.) \cr
	$|C_b|_2$ (fm$^{-1/2}$) & $3.2\times 10^4$ & $10\times 10^4$ \cr 
	\hline 
	$h_R^{(2)}\times 10^{-11}$ (MeV$^4$) 
	& $50.6\pm 0.4$ & 45.53$^{+0.04}_{-0.03}$ \cr
	$y^{(0)}$ (MeV$^{-1/2}$) &  
	$1.99\pm 0.01 \times 10^{-3}$ & 
	5.8$\pm 0.1\times 10^{-2}$ \cr \hline
	$\chi^2/N$ ($N=51$) & 1.55 & 1.18 \cr 
	$S_{E2}$ (keVb) at $E_G$ 
	& $4.1\pm 0.2$ & 
	42$^{+14}_{-13}$ \cr
	\hline
\end{tabular}
\caption{
	Values of $h_R^{(2)}$ and $y^{(0)}$ fitted to the experimental
	data of the $S_{E2}$ factor by using the two sets of values of the 
	effective range parameters, $r_2$, $P_2$, $Q_2$. 
	For the values in the second column of the table, the values
	of $r_2$, $P_2$, $Q_2$ presented in the third column of Table 
	\ref{table;parameters} are used; 
	the ANC of the $2_1^+$ state of $^{16}$O 
	is $|C_b|_2=3.2\times 10^{4}$~fm$^{-1/2}$. 
	For those in the third column of the table, we adjusted the values
	of $r_2$, $P_2$, $Q_2$ in the fourth column of 
	Table \ref{table;parameters} to reproduce the ANC deduced from 
	the $\alpha$-transfer reaction, $|C_b|_2 = 10\times 10^4$~fm$^{-1/2}$.
	$\chi^2/N$ values for fit are displayed in the table as well.  
	$S_{E2}$ at $E_G=0.3$~MeV 
	is calculated by using the fitted parameters.
}
\label{table;SE2}
\end{center}
\end{table}
In Table~\ref{table;SE2}, 
fitted values of the parameters, $h_R^{(2)}$ and $y^{(0)}$,
are presented with the $\chi^2/N$ values.
When fitting the parameters,
the values of the effective range parameters,
$r_2$, $P_2$, $Q_2$, displayed in the third column 
in Table~\ref{table;parameters} and in Eq.~(\ref{eq;r2P2Q2}) are used. 
Values of the $S_{E2}$ factor at $E_G=0.3$~MeV 
are calculated 
by using the fitted values of the parameters 
and displayed in the table as well. 
One can see in the table, the fitted values of the parameters, $h_R^{(2)}$
and $y^{(0)}$, are still scattered for the two cases.
The $\chi^2/N$ values in the last two columns
are $\chi^2/N= 1.55$ and 1.18, and the deduced values of $S_{E2}$ at $E_G$ 
show a difference about a factor of ten.
We have $S_{E2} = 4.1 \pm 0.2$ and $42^{+14}_{-13}$~keVb, respectively, 
where they have large, mostly about 33\% error bars. 
Those two values are still within the range of previously reported
values of the $S_{E2}$ factor summarized 
in Table IV in Ref.~\cite{detal-rmp17}. 

\begin{figure}
\begin{center}
  \includegraphics[width=13cm]{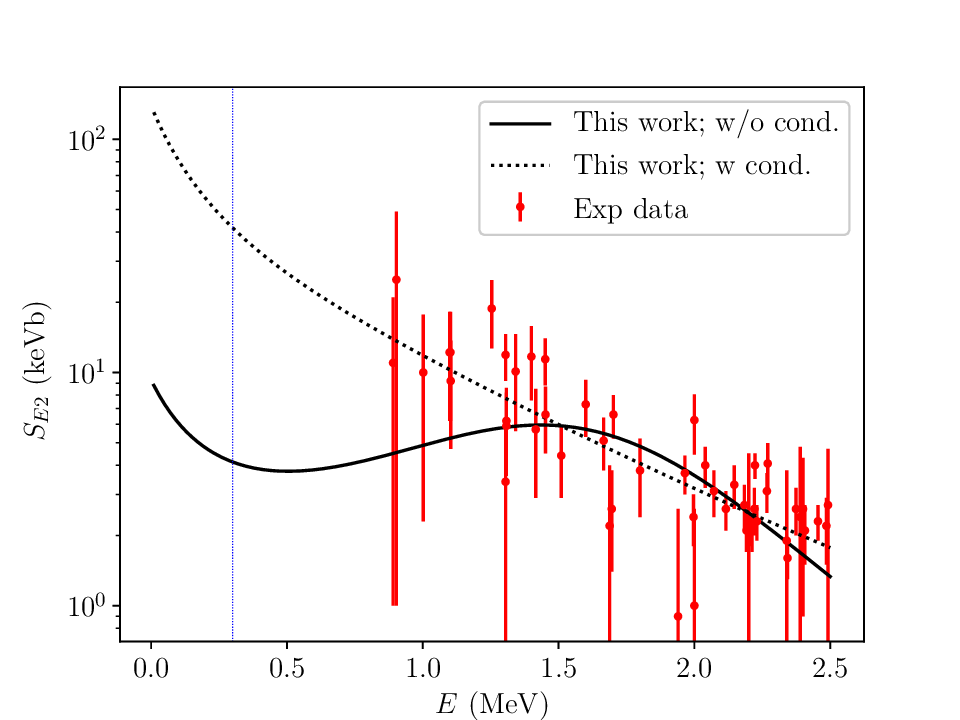}
	\caption{$S_{E2}$ factor of $^{12}$C($\alpha$,$\gamma$)$^{16}$O 
	plotted as functions of the energy $E$ of 
	the initial $\alpha$-$^{12}$C state in the center-of-mass frame.
	The two lines are plotted by using the fitted parameters 
	presented in Table \ref{table;SE2}.
	The experimental data are included in the figure as well. 
	The vertical blue line is drawn at $E_G=0.3$~MeV. 
}
\label{fig;SE2}       
\end{center}
\end{figure}
In Fig.~\ref{fig;SE2}, two lines of the $S_{E2}$ factor of 
$^{12}$C($\alpha$,$\gamma$)$^{16}$O are plotted 
as functions of the energy $E$
of the initial $\alpha$-$^{12}$C state in the center-of-mass frame.
The experimental data of the $S_{E2}$ factor are included in the figure as well.
A solid line of the $S_{E2}$ factor is calculated by using the fitted values
of the parameters to which the conditions to the effective range parameters
are not applied, and a dotted line of the $S_{E2}$ factor is by using those to
which the conditions are applied and one of the three effective range
parameters is constrained by the value of the ANC, $|C_b|_2 = 10\times 
10^4$~fm$^{-1/2}$, with Eq.~(\ref{eq;ANCnr2}).
One can see that 
the energy dependence of the $S_{E2}$ factor 
stems mainly from that of $D_2(E)$, 
which appear in the denominator
of the $E2$ transition amplitudes of $^{12}$C($\alpha$,$\gamma$)$^{16}$O,
displayed in Fig.~\ref{fig;ReD2}. 
The $\chi^2/N$ values of the two lines are 1.55 and 1.18, respectively,
and it indicates that the dotted line is better to fit the data than the 
solid line. 

\vskip 2mm \noindent
{\bf 4. Results and discussion}

In this work, we first studied 
the elastic $\alpha$-$^{12}$C scattering for $l=2$ introducing 
the conditions applied to the effective range 
parameters, $r_2$, $P_2$, $Q_2$, when fixing them to the phase shift 
data.
We employed the two data sets of the elastic scattering; 
one is precise phase shift data up to
the $p$-$^{15}$N breakup energy, $E_\alpha=6.62$~MeV, reported by Tischhauser
et al. (2009)~\cite{tetal-prc09}, and the other is the data set up to 
$E_\alpha=10$~MeV where the resonant $2_4^+$ state of $^{16}$O is covered
by the data, reported by Bruno et al. (1975)~\cite{betal-nc75}. 
We fit the parameters of the $S$ matrix of the elastic $\alpha$-$^{12}$C 
scattering for $l=2$ to the phase shift data 
for the two cases with and without the conditions applying 
to the effective range parameters 
in the low-energy region where no experimental
data are reported. 
We found the larger values of width of the $2_4^+$ state of 
$^{16}$O, $\Gamma_{R(24)}= 235(20)$ and  
$300^{+60}_{-40}$~keV, than that 
listed in the compilation, $\Gamma_{R(24)}=150(10)$~keV~\cite{twc-npa93},
and the large and small values of ANC of the $2_1^+$ state of $^{16}$O,
$|C_b|_2 = 23.3\times 10^4$ 
and $3.24\times 10^4$~fm$^{-1/2}$, 
for the two cases, respectively,
even though the two sets of the fitted parameters equally
reproduce the phase shift data well. 
The fitted values of the effective range parameters for the two cases were
applied to the study of the $S_{E2}$ factor 
of $^{12}$C($\alpha$,$\gamma$)$^{16}$O.
First, we study the energy dependence of the inverse of $^{16}$O propagator
for $l=2$ 
in the low energy region where the $S_{E2}$ factor is extrapolated to $E_G$.
Then, we fit the additional two parameters, 
$h_R^{(2)}$ and $y^{(0)}$, of the $E2$
transition amplitude of $^{12}$C($\alpha$,$\gamma$)$^{16}$O to the experimental
data of the $S_{E2}$ factor with the $\chi^2/N$ values, 
$\chi^2/N=1.18$ and 1.55, respectively, 
and extrapolate the $S_{E2}$ factor to $E_G$, where we have adjusted 
the effective range parameters for the case of the large ANC to 
reproduce the ANC of the $2_1^+$ state of $^{16}$O deduced from the 
$\alpha$-transfer reactions, $|C_b|_2=10\times 10^4$~fm$^{-1/2}$. 
We obtain $S_{E2}= 42^{+14}_{-13}$
and $4.1 \pm 0.2$~keVb at 
$E_G=0.3$~MeV; we find that both the values are within the range of previously 
reported values of $S_{E2}$ at $E_G$ in the literature. 

There is no restriction on whether one should apply the conditions to 
the effective range parameters or not when fitting to the phase shift data
because the phase shift data are equally well-fitted for both cases.
In other words, the phase shift data for $l=2$ cannot determine which line 
drawn in Fig.~\ref{fig;ReD2} is better than the other, while it is crucial
to extrapolate the $S_{E2}$ factor to $E_G$. 
One may argue a need to 
introduce the conditions employing an argument of 
the simplicity of natural laws,
as once discussed by Poincar\'e;  he wrote 
``natural laws must be simple''~\cite{p-1902}. 
For the present case, one may regard that the
dotted line (simply decreasing) is simpler than 
the solid line (having a plateau) in Fig.~\ref{fig;ReD2};
the appearance of such a bump of the $S_{E2}$ factor 
might indicate interference with an unknown bound or resonant state 
at the low energies. 
While such an assumption should be tested by experiment 
or other possible methods.

A quantity which could test a verification of the conditions may be the 
width of the resonant $2_4^+$ state of $^{16}$O. The reported values displayed 
in Table~\ref{table;E24rnG24r} are still scattered, but the 
value, $\Gamma_{R(24)}= 349(3)$~keV, recently reported by deBoer et al. (2012) 
could support the result of  
$\Gamma_{R(24)} = 300^{+60}_{-40}$~keV, 
which was obtained without applying the conditions.  
Meanwhile, as discussed above,
we need to improve the treatment in theory because the new channels start
opening in this energy region. 

The experimental data of 
the $S_{E2}$ factor of $^{12}$C($\alpha$,$\gamma$)$^{16}$O
may provide another quantity to test verification of the conditions
because the data cover the lower energy region, $E=0.9$ -- $1.95$~MeV 
($E_\alpha = \frac43 E=1.2$ -- $2.6$~MeV) than those of the elastic 
$\alpha$-$^{12}$C scattering though 
the data of the $S_{E2}$ factor  
have large error bars, especially in the lower energy region, 
$E=0.9$ -- $1.2$~MeV.
After fitting the two parameters, $h_R^{(2)}$ and
$y^{(0)}$, of the $E2$ transition amplitudes of 
$^{12}$C($\alpha$,$\gamma$)$^{16}$O, we have the $\chi^2/N$ values for the
two cases as $\chi^2/N=1.18$ and 1.55; 
this may support to apply the conditions in the low-energy region 
while the data of the $S_{E2}$ factor still have a large uncertainty. 
More accurate measurements of the $S_{E2}$ factor in the energy range,
$E=0.9$ -- $1.5$~MeV, would be helpful to obtain a clear conclusion.

The values of the ANC 
of the $2_1^+$ state of $^{16}$O we obtained in this work are still 
quite different for the two cases, $|C_b|_2 = 3.24\times 10^4$ and 
$23.3\times 10^4$~fm$^{-1/2}$. As mentioned, 
the values of $|C_b|_2$ are deduced from 
the $\alpha$ transfer reactions, 
such as $^{12}$C($^6$Li,$d$)$^{16}$O~\cite{aetal-prl15}; 
the value of $|C_b|_2$ is recently updated by Hebborn et al. as 
$|C_b|_2 = 10.7(3)\times 10^4$~fm$^{-1/2}$~\cite{hetal-23} 
by using the ANC of the ground state of $^6$Li as $d$-$\alpha$ system 
deduced from their ab initio calculation~\cite{hetal-prl22}. 
As discussed above, we have employed a value of the ANC,
$|C_b|_2=10\times 10^4$~fm$^{-1/2}$, deduced from the $\alpha$-transfer
reactions to constrain the values of the effective range parameters
by Eq.~(\ref{eq;ANCnr2}) when fitting them to the phase shift data
applying the conditions,
and we have $S_{E2}=42^{+14}_{-13}$~keV\,b with $\chi^2/N=1.18$. 
The $\chi^2$ value is small but the error of the $S_{E2}$ factor is 
significantly large, about 33\% error. This may also stem from the 
large errors of the data of the $S_{E2}$ factor.
Thus, it would be important to reduce the error of the $S_{E2}$ factor
by using the other experimental data. 
The study in this direction is now under investigation. 

\vskip 2mm \noindent
{\bf Acknowledgements}

The author would like to thank 
D. Phillips, 
T. Kajino, R. deBoer, 
and 
C.~H. Hyun
for discussions. 
This work was supported by
the National Research Foundation of Korea (NRF) grant funded by the
Korean government (MSIT) (No. 2019R1F1A1040362 and 2022R1F1A1070060)
and the Korean Evaluation Institute of Industrial Technology (KEIT) 
grand funded by the Korean government (MOTIE) (No. 20022473). 

\vskip 2mm \noindent
{\bf Appendix A}

In this appendix, we discuss the relations related to the function $H_2(p)$
in Eq.~(\ref{eq;H2}) in the low energy limit, $p\to 0$.
Using two formulas of the digamma function $\psi(z)$;
one is Eq.~6.3.18 in Ref.~\cite{as14}, 
\bea
\psi(z) &\sim& \ln z 
- \frac{1}{2z} 
- \sum_{n=1}^\infty \frac{B_{2n}}{2nz^{2n}}\,
\nnb \\ &=&
\ln z
- \frac{1}{2z} 
- \frac{1}{12 z^2}
- \frac{1}{120 z^4}
- \frac{1}{252 z^6}
- \cdots\,,
\eea
for $|z|\to \infty$ and $|arg z| <\pi$, where $B_{2n}$ are the Bernoulli
numbers, 
\bea
B_2 = \frac16\,, \ \ 
B_4 = - \frac{1}{30}\,, \ \ 
B_6 = \frac{1}{42}\,, \ \ 
B_8 = - \frac{1}{30}\,, \ \ 
B_{10} = \frac{5}{66}\,,  \ \ 
\cdots\,,
\eea
and the other is Eq.~5.4.16 in Ref.~\cite{olbc10},
\bea
Im \psi(iy) = \frac{1}{2y} + \frac{\pi}{2}\coth(\pi y)\,,
\eea
one can rewrite the imaginary part and real part of the function $H(\eta)$
in Eq.~(\ref{eq;H2}) as
\bea
Im H(\eta) &=& Im \psi(i\eta) - \frac{1}{2\eta} -\pi 
= \frac{1}{2\eta} \frac{2\pi \eta}{e^{2\pi\eta}-1}
= \frac{1}{2\eta}C_\eta^2\,,
\\ 
Re H(\eta) &=& Re \psi(i\eta) - \ln\eta
= - \sum_{n=1}^\infty \frac{B_{2n}}{2n(i\eta)^{2n}}
\nnb \\ &=&
\frac{1}{12\eta^2}
+ \frac{1}{120\eta^4}
+ \frac{1}{252\eta^6}
+ \frac{1}{240\eta^8}
+ \frac{1}{132\eta^{10}}
+ \cdots\,.
\eea
Now one may obtain the expression of $2\kappa Re H_2(p)$ in 
Eq.~(\ref{eq;2kappaReH2}).

The expressions of $ReD_2(p)$ in Eq.~(\ref{eq;ReD2}) is calculated 
as the following.
First one may expand $H(\eta)$ function using the equation above, and one
has an expression of $ReD_2(p)$ as
\bea
ReD_2(p) = a(\gamma_2^2 + p^2) 
+ b (\gamma_2^4 - p^4) 
+ c (\gamma_2^6 + p^6)
+ d (\gamma_2^8 - p^8) 
+ e (\gamma_2^{10} + p^{10})
+ \cdots\,,
\eea
where $a$, $b$, $c$, $d$, $e$ are coefficients. 
Explicitly, we have
\bea
ReD_2(p) &=& \left(\frac12 r_2 - \frac{1}{24}\kappa^3\right)(\gamma_2^2+p^2)
+\left(
\frac14P_2 + \frac{17}{80}\kappa
\right)(\gamma_2^4 - p^4)
\nnb \\ &&
+ \left(
Q_2 - \frac{757}{4032\kappa}
\right)(\gamma_2^6+p^6)
+ \frac{289}{10080\kappa^3}(\gamma_2^8 -p^8)
- \frac{491}{22176\kappa^5}(\gamma_2^{10} + p^{10})
\nnb \\ &&
+\cdots\,.
\eea
Then, one may use the relations,
\bea
\gamma_2^4 - p^4 &=& - (\gamma_2^2 + p^2)^2 + 2\gamma_2^2 (\gamma_2^2 + p^2)\,,
\\
\gamma_2^6 + p^6 &=& (\gamma_2^2 + p^2)^3 -3\gamma_2^2 (\gamma_2^2+p^2)^2
+ 3\gamma_2^4(\gamma_2^2+p^2)\,,
\\
\gamma_2^8 -p^8 &=& -(\gamma_2^2+p^2)^4 
+ 4 \gamma_2^2 (\gamma_2^2 + p^2)^3 
- 6 \gamma_2^4 (\gamma_2^2+p^2)^2
+ 4 \gamma_2^6 (\gamma_2^2 + p^2) \,,
\\
\gamma_2^{10} + p^{10} &=& (\gamma_2^2 + p^2)^5
-5\gamma_2^2(\gamma_2^2 + p^2)^4
+10\gamma_2^4(\gamma_2^2+p^2)^3
-10\gamma_2^6(\gamma_2^2+p^2)^2
\nnb \\ &&
+5\gamma_2^8(\gamma_2^2+p^2)\,,
\eea
and has the expression
\bea
ReD_2(p) &\simeq& \sum_{n=1}^5 C_n (\gamma_2^2 + p^2)^n\,,
\eea
with
\bea
C_1 &=& a + 2\gamma_2^2 b + 3\gamma_2^4 c + 4 \gamma_2^6 d + 5 \gamma_2^8 e\,,
\\
C_2 &=& -b -3\gamma_2^2 c - 6 \gamma_2^4d -10\gamma_2^6 e\,,
\\
C_3 &=& c + 4\gamma_2^2 d + 10\gamma_2^4 e\,,
\\
C_4 &=& -d - 5\gamma_2^2 e\,,
\\
C_5 &=& e\,.
\eea
Then, one may obtain the expressions of the coefficients,
$C_i$ with $i=1,2,3,4,5$
in Eqs.~(\ref{eq;C1},\ref{eq;C2},\ref{eq;C3},\ref{eq;C4},\ref{eq;C5}). 

\vskip 2mm \noindent
{\bf Appendix B}

In this appendix, we present the expression of the $E2$ transition amplitudes
of $^{12}$C($\alpha$,$\gamma$)$^{16}$O and briefly discuss its 
derivation. 
\begin{figure}[t]
\begin{center}
\includegraphics[width=10cm]{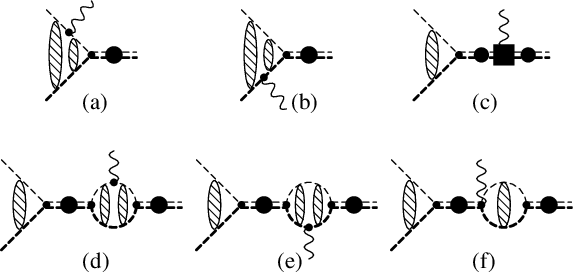}
\caption{
	Diagrams of amplitudes for radiative $\alpha$ capture on $^{12}$C.
	A wavy line and a thin (thick) dashed line denote the outgoing photon
	and $\alpha$ ($^{12}$C) state, respectively. Double thin-and-thick
	lines with a filled circle denote the dressed $^{16}$O 
	propagators for $2_1^+$ state in the intermediate state and 
	for $0_1^+$ state in the final state. 
	See the caption in Fig.~\ref{fig;propagator} as well.
	A $O\gamma O^*$ vertex in the diagram (c) is a counter term 
	proportional to $h_R^{(2)}$
	to renormalize the infinities from loop diagrams in (d), (e), (f).  
}
\label{fig;e2-amplitudes}
\end{center}
\end{figure}
In Fig.~\ref{fig;e2-amplitudes}, the diagrams of the reaction are displayed.
The vertex functions and propagators are derived from the effective Lagrangian
and 
we have the $E2$ transition amplitude of $^{12}$C($\alpha$,$\gamma$)$^{16}$O 
as
\bea
A^{(l=2)} &=& \vec{\epsilon}_{(\gamma)}^* \cdot \hat{p}
\hat{k}'\cdot\hat{p} X^{(l=2)}\,,
\eea
where $\vec{\epsilon}^*_{(\gamma)}$ 
is the polarization vector of the outgoing photon,
$\hat{k}'$ is the unit vector of photon three-momentum, and $\hat{p}$ 
is the unit vector of relative momentum of the initial $\alpha$-$^{12}$C 
system. The amplitude $X^{(l=2)}$ is decomposed as 
\bea
X^{(l=2)} &=&
X^{(l=2)}_{(a+b)}
+ X^{(l=2)}_{(c)}
+ X^{(l=2)}_{(d+e)}
+ X^{(l=2)}_{(f)}
\,,
\eea
where each amplitude corresponds to the diagrams 
depicted in Fig.~\ref{fig;e2-amplitudes}. 
Thus, we have
\bea
X_{(a+b)}^{(l=2)} &=&
- 6y^{(0)} e^{i\sigma_2} \Gamma(1+\kappa/\gamma_0)
\nnb \\ && \times
\int_0^\infty drrW_{-\kappa/\gamma_0,\frac12}(2\gamma_0 r)
\left[
\frac{Z_\alpha \mu}{m_\alpha}j_1\left(
\frac{\mu}{m_\alpha} k'r
\right)
+\frac{Z_C \mu}{m_C}j_1\left(
\frac{\mu}{m_C} k'r
\right)
\right]
\nnb \\ && \times
\left(
\frac{\partial}{\partial r} + \frac{3}{r}
\right)
\frac{F_2(\eta,pr)}{pr}\,,
\\
X_{(c)}^{(l=2)} &=&
+ y^{(0)} 
\left\{
	- h_R^{(2)} 
	+ \frac{3\kappa\mu^3m_O^2}{2\pi Z_O} 
	\left( 
	\frac{Z_\alpha}{m_\alpha^2} + \frac{Z_C}{m_C^2}
	\right)
	\left[
		\frac{4}{225}
	\ln\left(
	\frac{\mu_{DR}}{2}r_C
	\right)
	- \ln\left(
	\frac{\mu_{DR}}{\kappa}
	\right)
	\right]
	\right\}
\nnb \\ && \times
\frac{5\pi Z_O}{\mu m_O^2}
\frac{e^{i\sigma_2}k'p^2\sqrt{(1+\eta^2)(4+\eta^2)}C_\eta}{
	K_2(p) - 2\kappa H_2(p)}\,,
\label{eq;Xc}
\\
X_{(d+e)}^{(l=2)} &=&
+ \frac{1}{5}y^{(0)}
\frac{
e^{i\sigma_2}p^4\sqrt{(1+\eta^2)(4+\eta^2)}C_\eta
}{
	K_2(p)-2\kappa H_2(p)
}
\Gamma(1+\kappa/\gamma_0)
\Gamma(3+i\eta)
\nnb \\ && \times
\int_{r_C}^\infty drrW_{-\kappa/\gamma_0,\frac12}(2\gamma_0 r)
\left[
\frac{Z_\alpha \mu}{m_\alpha}j_1\left(
\frac{\mu}{m_\alpha} k'r
\right)
+\frac{Z_C \mu}{m_C}j_1\left(
\frac{\mu}{m_C} k'r
\right)
\right]
\nnb \\ && \times
\left(
\frac{\partial}{\partial r} + \frac{3}{r}
\right)
\frac{W_{-i\eta,\frac52}(-2ipr)}{r}\,,
\label{eq;Xde}
\\
X_{(f)}^{(l=2)} &=& - \frac{15}{4} y^{(0)} \mu^2\left(
\frac{Z_\alpha}{m_\alpha^2} + \frac{Z_C}{m_C^2} 
\right) \left[-2\kappa H(\eta_{0b})\right]
\frac{e^{i\sigma_2} k'p^2\sqrt{(1+\eta^2)(4+\eta^2)}C_\eta}{
	K_2(p) - 2\kappa H_2(p)
}
\,,
\eea
where $m_\alpha$, $m_C$, $m_O$ ($Z_\alpha$, $Z_C$, $Z_O$) are the mass of
(the number of protons in) $\alpha$, $^{12}$C, $^{16}$O, respectively. 
$\mu$ and $\kappa$ are the reduced mass and the inverse of the Boer radius
of $\alpha$-$^{12}$C system.
$k'$ and $p$ are the magnitudes of three momentum of outgoing photon
and that of relative momentum of the $\alpha$-$^{12}$C system 
in the center-of-mass frame. 
$\eta$ is the Sommerfeld parameter $\eta = \kappa/p$. 
$\gamma_0$ is the binding momentum of the ground state of $^{16}$O; 
$\gamma_0 = \sqrt{2\mu B_0}$ where $B_0$ is the binding energy 
of $\alpha$-$^{12}$C system in the ground state of $^{16}$O, and  
$\eta_{0b} = \kappa/(i\gamma_0)$.
$\Gamma(z)$, $j_l(x)$, $F_l(\eta,\rho)$, $W_{\kappa,\mu}(z)$ 
are the gamma function,
the spherical Bessel function, the regular Coulomb function, 
and the Whittaker function, respectively. 
$\sigma_2$ is the Coulomb phase shift for $l=2$.

The three loop diagrams of the $O\gamma O^*$ vertex in the 
figures (d), (e), (f) in Fig.~\ref{fig;e2-amplitudes} diverge.
The log divergence appears in the $r$-space integral in $r\to 0 $ limit
in Eq.~(\ref{eq;Xde}) for the diagrams (d) and (e); 
we introduce a cutoff $r_C$ in the $r$-space integral and the infinite part
is renormalized by the counter term, $h^{(2)}$, in Eq.~(\ref{eq;Xc}).
The divergence appearing in the diagram (f) was regulated in the 
momentum space integral as $J_0^{div}$ by means of the 
dimensional regularization~\cite{kr-plb99,setal-prc07}. 
Those infinities are
renormalized by the renormalized coefficient, $h_R^{(2)}$, as 
\bea
\lefteqn{
- h^{(2)} 
+ \frac{3\mu^2 m_O^2}{2Z_O}
\left(
\frac{Z_\alpha}{m_\alpha^2} + \frac{Z_C}{m_C^2}
\right)
\left[
      -	J_0^{div} + \frac{4\kappa\mu}{225\pi}
\left(
\frac{\mu_{DR}}{2} 
\right)^{2\epsilon} 
\int_0^{r_C}\frac{dr}{r^{1-2\epsilon}}
\right]
}
\nnb \\ &=& 
- h_R^{(2)} 
+ \frac{3\kappa\mu^3m_O^2}{2\pi Z_O}
\left(
\frac{Z_\alpha}{m_\alpha^2} + \frac{Z_C}{m_C^2}
\right)
\left[
- \ln\left( \frac{\mu_{DR}}{\kappa}
\right)
+ 	\frac{4}{225}
\ln\left(
\frac{\mu_{DR}}{2}r_C
\right)
+ O(\epsilon) 
\right]\,,
\eea
with
\bea
J_0^{div} &=&\frac{\mu\kappa}{2\pi}\left[
	\frac{1}{\epsilon} - 3C_E + 2 + \ln\left(
	\frac{\pi\mu_{DR}^2}{4\kappa^2}
	\right)
	\right]\,,
\eea
where we performed the integration in $d=4-2\epsilon$ dimensions, 
and
$\mu_{RD}$ is the scale of the dimensional regularization
and $C_E$ is the Euler's constant, $C_E = 0.5771\cdots$; 
we choose $\mu_{DR} = \Lambda_H = 160$~MeV. 
We found that a minor cutoff $r_C$ dependence
and choose $r_C = 0.01$~fm. 
The $E2$ transition amplitudes up to this order have two additional parameters,
$h_R^{(2)}$ and $y^{(0)}$, along with the effective range parameters,
$r_2$, $P_2$, $Q_2$ in the function of $K_2(p)$. 

The total cross-section is
\bea
\sigma_{E2} &=& \frac43\frac{\alpha_E \mu E_\gamma'}{p
(1+E_\gamma'/m_O)} 
\frac15|X^{(l=2)}|^2 
\,,
\eea
where $E_\gamma' (=k')$ is the energy of outgoing photon, 
\bea
E_\gamma' &\simeq& B_0 + E - \frac{1}{2m_O}(B_0+E)^2\,,
\eea
and the $S_{E2}$ factor is defined as 
\bea
S_{E2}(E) &=& \sigma_{E2}(E)Ee^{2\pi\eta}\,.
\eea

\end{document}